# Immersive Virtual Reality Serious Games for Evacuation Training and Research: A Systematic Literature Review

Zhenan Feng, Vicente A. González, Robert Amor, Ruggiero Lovreglio, Guillermo Cabrera-Guerrero

## Abstract

An appropriate and safe behavior for exiting a facility is key to reducing injuries and increasing survival when facing an emergency evacuation in a building. Knowledge on the best evacuation practice is commonly delivered by traditional training approaches such as videos, posters, or evacuation drills, but they may become ineffective in terms of knowledge acquisition and retention. Serious games (SGs) are an innovative approach devoted to training and educating people in a gaming environment. Recently, increasing attention has been paid to immersive virtual reality (IVR)-based SGs for evacuation knowledge delivery and behavior assessment because they are highly engaging and promote greater cognitive learning.

This paper aims to understand the development and implementation of IVR SGs in the context of building evacuation training and research, applied to various indoor emergencies such as fire and earthquake. Thus, a conceptual framework for effective design and implementation through the systematic literature review method was developed. As a result, this framework integrates critical aspects and provides connections between them, including pedagogical and behavioral impacts, gaming environment development, and outcome and participation experience measures.

**Keywords:** Virtual Reality; Serious Games; Evacuation Study; Systematic Literature Review.

## 1. Introduction

Today, most people spend a large portion of their time living and working in buildings. However, natural or man-made hazards can make the building environment dangerous for staying. Proper evacuation responses and behavior during an emergency is a crucial factor to increase survival chance. In general, people are trained and acquire evacuation knowledge (e.g., emergency response, self-protection skills, best practice) through traditional approaches such as videos, posters, seminars, courses, or evacuation drills. However, these traditional approaches may not effectively transmit knowledge (Gwynne, Boyce, Kuligowski, Nilsson, P. Robbins, & Lovreglio, 2016). One reason is that after an evacuation drill, building occupants are generally not provided with individual feedback assessing their evacuation behavior (Gwynne, Kuligowski, Boyce, Nilsson, Robbins, Lovreglio et al., 2017). Another reason is that building occupants can be not emotionally engaged in the learning process that may lead to a reduced effect on attitude and limited change in behavior (Chittaro, Buttussi, & Zangrando, 2014). In fact, Yang et al. (2011) pointed out that real-life evacuation behavior is different from experiments such as evacuation drills, which means that current evacuation models still have limitations when they are the basis of evacuation training and research. Evacuation drills also have other limitations such as being costly in time and resources by interrupting building occupants' routines, and being not able to present hazards (Gwynne et al., 2016; Gwynne et al., 2017; Silva, Almeida, Rossetti, & Leca Coelho, 2013). Therefore, there is a need to investigate innovative and more effective approaches to overcome the limitations mentioned above (Kobes, Helsloot, de Vries, & Post, 2010). Such innovations should be able to transmit evacuation knowledge of building occupants towards one more effective and efficient.

In that regard, Serious Games (SGs) have attracted much attention on pedagogical research recently (Connolly, Boyle, Boyle, MacArthur, & Hainey, 2012). SGs are video games whose primary purposes are training and education, not entertainment per se (Connolly, Stansfield, & Boyle, 2009). It is argued that by playing SGs, participants can gain and retain knowledge more effectively than by using traditional learning methods (Wouters, Van Nimwegen, Van Oostendorp, & Van Der Spek, Eric D, 2013). SGs can enhance the capability of traditional approaches to deliver evacuation knowledge (Mayer, Wolff, & Wenzler, 2013; Zhou, Chang, Pan, & Whittinghill, 2016). In turn, there is another technology that can promote the engaging capabilities of SGs namely Immersive Virtual Reality (IVR). IVR allows participants to be fully immersed in virtual environments that can provide greater engagement and perception than videos, text-based papers or 2D games (Gao, Gonzalez, & Yiu, 2017; Lovreglio, Gonzalez, Amor, Spearpoint, Thomas, Trotter et al., 2017). The combination of IVR and SGs encourages participants to retain knowledge longer than traditional approaches due to the fact that they benefit from full engagement, and high emotional and physiological arousal (Chittaro & Buttussi, 2015). As such, an increasing number of studies have investigated the combination of IVR and SGs for evacuation training and behavior assessment. A systematic understanding of how IVR SGs have been developed and implemented for evacuation training and research is necessary.

This paper introduces a systematic literature review regarding IVR SGs oriented toward building evacuation processes tailored to indoor emergencies. As a result, a conceptual framework to guide the development and implementation of such IVR SGs is proposed. Thus, the key factors contributing to successful and comprehensive development and implementation of IVR SGs tailored to building evacuation are identified and connected.

## 2. Background

### 2.1 Serious Games

SGs have become a popular training and behavior analysis tool in the last decades (Connolly et al., 2012). The term "serious game" usually represents a video game whose primary purpose is education, training, simulation, socializing, exploring, analyzing and advertising, rather than pure entertainment (Michael & Chen, 2005). Susi et al. (2007) suggested that SGs represent "*the application of gaming technology, process, and design to the solution of problems faced by businesses and other organizations. SGs promote the transfer and cross-fertilization of game development knowledge and techniques in traditionally non-game markets such as training, product design, sales, marketing, etc.*" Note that SGs are not only able to transfer knowledge, but are also able to fulfil other objectives such as behavior analysis (Issa & Zhang, 2015) and rehabilitation healthcare (Schonauer, Pintaric, Kaufmann, Jansen - Kosterink, & Vollenbroek-Hutten, 2011).

One of the primary objectives of SGs is to educate participants (Connolly et al., 2012), and SGs have been investigated widely in different domains for education purposes. Connolly et al. (2012) suggested that SGs can be applied to acquiring knowledge, understanding content, or developing specific skills. For instance, Johnson and Wu (2008) explored the capability of SGs to facilitate teaching foreign languages. Muratet et al. (2009) proposed an SG to improve programming skills, and Sliney and Murphy (2008) and Diehl et al. (2011) developed SG prototypes for medical training. There has also been a number of studies focused on emergency training. SGs have been implemented for emergency training in oil industry (Mayer et al., 2013; Metello, Casanova, & Carvalho, 2008), terrorist attacks (Chittaro & Sioni, 2015), fire evacuation (Chittaro & Ranon, 2009; Sacfung, Sookhanaphibarn, & Choensawat, 2014), disaster evacuation (Cohen, Sevdalis, Taylor, Kerr, Heys, Willett et al., 2012), and earthquake evacuation (Barreto, Prada, Santos, Ferreira, O'Neill, & Oliveira, 2014; Tanes & Cho, 2013). All these studies suggested that SGs are a promising tool for education and training purposes. One explanation is that participants can recall more effectively what they have learned compared to traditional learning approaches (Bartolome, Zorrilla, & Zapirain, 2011). Papastergiou (2009) argued that SGs could positively motivate participants during the learning experience. Another explanation is that participants have the chance to interact with the environment and get immediate feedback from the SGs to rectify any incorrect responses and to strengthen knowledge (Lovreglio, Gonzalez, Feng, Amor, Spearpoint, Thomas et al., 2018). SGs are an effective tool to support and enhance traditional training tools (Gao, Gonzalez, & Yiu, 2018).

Another objective of SGs is to investigate human behavior (T. M. Connolly et al., 2012). Their gaming structure enables tracking and recording participants' decisions and behavior during a game experience. By collecting and analyzing behavioral data, it is possible to reveal behavioral motivation, validate behavioral models, explore decision-making, recognize behavioral patterns, and assess the responses under various controlled conditions (Connolly et al., 2012). For instance, one study (Chittaro & Ranon, 2009) adopted VU-Flow tool (VU-Flow provides a set of interactive visualizations that highlight interesting navigation behavior of single or groups of moving entities that were in the virtual environment together or separately) with the game to track and visualize participants' fire evacuation routes, in order to understand the evacuation navigation patterns. In another study (Li, Liang, Quigley, Zhao, & Yu, 2017), participants' awareness of potential hazards during an earthquake was simulated by tracking visual attention to falling and fragile objects inside the game environment. Therefore, SGs have potential to allow the understanding of behavioral patterns and behavior changes beyond educational and training aspects.

## 2.2 Immersive Virtual Reality

Virtual Reality (VR) technology has rapidly evolved in recent years, bringing a wide range of application areas due to its flexibility to adapt to different problems and domains (LaValle, 2017). This has also brought different interpretations of what VR is. In this paper, we refer to VR to that experience in which participants are fully immersed into the virtual environment provided by head-mounted displays (HMD) or projection-based displays (PBD) (Sharples, Cobb, Moody, & Wilson, 2008). The most popular PBD application is the CAVE system (cave automatic virtual environment), which displays images on screens formed up a 3D immersive room, providing an immersive visual and interaction experience (Cruz-Neira, Sandin, & DeFanti, 1993). Figure 1(a) shows the HMD system and Figure 1(b) shows the PBD system.

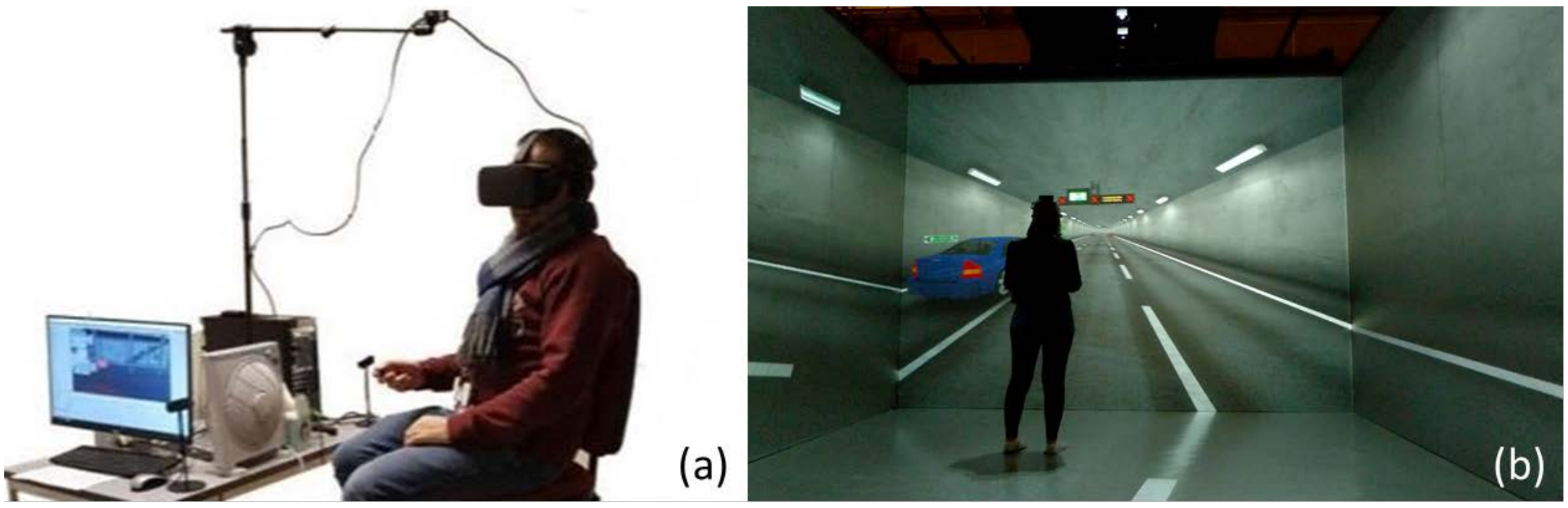

Figure 1 – (a) HMD system; (b) PBD system

We use the term of “immersive virtual reality (IVR)” to distinguish it from other forms of VR. IVR can be defined as “*Inducing targeted behavior in an organism by using artificial sensory stimulation, while the organism has little or no awareness of the interference*” (LaValle, 2017). According to this definition, once participants are immersed into the virtual environment, they can feel they are physically inside this environment, even though such environment is artificially simulated. This virtual environment may become very realistic, making it very difficult for individuals to differentiate between the virtual and the real world (LaValle, 2017). Therefore, IVR has the potential to allow participants to behave and react as close as possible to reality (Sherman & Craig, 2003).

IVR can create a number of benefits resulting from its applications. IVR has been demonstrably effective in communicating the cultural content of a museum exhibition (Carrozzino & Bergamasco, 2010). It also can perform social interaction through behavior and context to improve the learning environment (Bailenson, Yee, Blascovich, Beall, Lundblad, & Jin, 2008). In another study (Blascovich, Loomis, Beall, Swinth, Hoyt, & Bailenson, 2002), IVR was proposed as a social psychological research tool to lessen the trade-off between mundane realism and experimental control. IVR is also an ideal tool for exploration, training, and education (Psotka, 1995). As a result, its potential for emergency training and education has been investigated (Li et al., 2017; Sharma, Jerripothula, Mackey, & Soumare, 2014). Shendarkar et al. (2008) suggested that it is more effective to conduct emergency management by using IVR due to its capability to model human behavior with a high degree of fidelity. Smith and Ericson (2009) revealed that IVR could enhance the enthusiasm of children for fire-safety skills training by improving their engagement with the learning environment.

A recent study by Krokos et al. (2018) found that IVR can provide better memory recall ability compared to non-IVR conditions. Their research showed that participants felt more focused on the task resulting from better immersion experience. In addition, the majority of the participants also claimed that the sense of the spatial awareness enhanced by IVR was critical to their success. Given that, IVR experience can influence SGs elements to make the learning and behavior outcome significant resulting from its special cognitive process. SGs can be enhanced for training and education purposes using IVR principles, and thus, they can be regarded as IVR SGs. The involvement of IVR principles in SGs provides a higher degree of engagement when compared to non-IVR SGs (Gao et al., 2017; Lovreglio et al., 2017). Chittaro and Buttussi (2015) suggested that IVR SGs can improve knowledge retention and psychological arousal in aviation safety training. One possible interpretation is that IVR SG frameworks have the capability to generate “realistic scenarios” and hazards, which in turn create more “realistic feelings” in participants. The effect is to make evacuation simulation highly engaging so that participants can get better hazards perception and risk awareness. Apart from the benefit of training and education, IVR SGs are also valuable tools for analyzing human behavior during different emergencies such as fire (Kinateder, Ronchi, Nilsson, Kobes, Muller, Pauli et al., 2014) or earthquakes (Li et al., 2017). Kinateder et al. (2014) argued that IVR SGs allow a safe study of occupant behavior in scenarios that would be too dangerous to implement in the real world evacuation drills such as dense smoke or falling objects.

However, IVR SGs have limitations. One significant issue is that a VR environment may induce motion sickness (Hettinger & Riccio, 1992). Sharples et al. (2008) argued that there is a high chance of producing virtual reality induced side effects (motion sickness) by using head-mounted displays. A possible explanation for this side effect is that participants suffer from sensory conflicts when they view compelling visual representations of self-motion in the IVR environment with physically stationary self-body (Hettinger & Riccio, 1992). Regan and Price (1994) stated that 61% participants felt symptoms of malaise such as dizziness, nausea, and headache during the simulation. However, Lovreglio et al. (2018) reported that only 5% participants felt motion sickness in their study, that was benefited from the high quality of navigation systems. Therefore, a thoughtful design of IVR SGs is necessary to minimize these side effects.

Various applications of IVR have been explored by researchers. However, so far there has been no literature review systematically assessing the combination of IVR and SGs for building evacuation purposes. Given that, there is still a need to understand how to systematically develop and implement IVR SGs for building evacuation training and research.

## 3. Research Design

Inspired by the capabilities and potential of IVR, this research is focusing on the combination of IVR and SGs. This research aims to provide insight about the development and implementation criteria of IVR SGs oriented towards indoor evacuation processes. A conceptual framework is expected to be generated as the main outcome of this research.

In order to achieve this objective, a systematic literature review was conducted to comprehensively explore the existing IVR SGs tailored to building evacuation training and research. A qualitative data analysis was carried out to identify the empirical evidence on the essential factors contributing to an effective and robust development and implementation of IVR SGs suited to indoor evacuation training and research. As a result, a framework was generated integrating the analyzed evidence providing guidelines for future research.

## 4. Systematic Literature Review

The systematic literature review was conducted in accordance with the framework recommended by Khan et al. (2003) and Thomé et al. (2016). This review included five stages: formulating the research problems, identifying relevant work, assessing the quality of studies, summarizing the evidence, and interpreting the findings.

### 4.1 Formulating the Research Problems

Rüppel and Schatz (2011) elaborated a triadic game design approach to SGs that included three interdependent worlds that need to be balanced during the design process: reality (how the game is connected to the physical world), meaning (what value needs to be achieved), and play (how to create playful activities). We investigated these three major aspects for our research. The investigation allowed us to answer two main research questions. In order to get a detailed understanding of these two main questions, eleven sub-questions were formulated. Table 1 shows the question, sub-questions and assessed aspects.

Table 1

Systematic literature review research questions.

| Main research questions | Sub-questions | Assessed aspects |
|---|---|---|
| MQ1: What are the outcomes and measures for implementing IVR SGs in evacuation study? | SQ1: What are the learning outcomes? | Meaning: Pedagogical impact |
| | SQ2: What are the learning measures? | |
| | SQ3: What are the behavior outcomes? | Meaning: Behavioral impact |
| | SQ4: What are the behavior measures? | |
| | SQ5: How to evaluate the participation experience? | Play: Participation experience |
| MQ2: What are the essential elements for developing IVR SGs in evacuation study? | SQ6: What are the teaching methods? | Play: Hardware and software system |
| | SQ7: What are the navigation solutions? | |
| | SQ8: What are the senses stimulated? | |
| | SQ9: What are the narrative methods to encourage the participants to follow the game storyline, and complete it? | |
| | SQ10: Are there non-playable characters (NPCs) and how do they contribute? | |
| | SQ11: What are the hazards simulated? | Reality: Software system |

In general, two types of impacts of IVR SGs for evacuation training and research can be summarized; one is a pedagogical impact, and the other is a behavioral impact (Lovreglio et al., 2018). SQ1 and SQ2 were formulated to explore the pedagogical impact, while SQ3 and SQ4 were for the behavioral impact. Apart from that, participation experience is also an important aspect that can be used to support and refine the prototype (Chittaro & Buttussi, 2015). On that basis, SQ5 was formulated to explore how to measure the participation experience. In terms of the gaming environment development, it includes various components, which could be either the hardware system and software system (Rüppel & Schatz, 2011). Therefore, RQ6 to RQ11 were formulated to discover the specific details of different systems, including the teaching methods, navigation solutions, sense stimulation, narrative methods, NPCs, and hazards simulation.

## 4.2 Identifying the Relevant Work

The eligible papers need to include three major concepts, namely, immersive virtual reality, serious games, and evacuation training and research. IVR is a mechanism added onto SGs, while SGs are still the key content in the eligible papers which cover existing knowledge on SGs to achieve their primary research aims.

Eligible papers included in the systematic literature review were collected from journals and conference proceedings. The papers were recovered from the following databases: Scopus and Engineering Village. Scopus is the largest abstract and citation database of peer-reviewed literature, and Engineering Village is an index of the most comprehensive engineering literature. Meanwhile, another approach called snowballing (retrieve relevant papers based on target papers' references list or paper citing) (Wohlin, 2014, May) was also adopted with Google Scholar, which indexes most scholarly literature, as a complementary method to cover any missing papers.

There is an inconsistency in the terminology used in the literature. For instance, virtual reality, virtual environment, virtual simulation or VR can all represent the content of immersive virtual reality. To get the maximum coverage of publications, we conducted searches using the following search string: "virtual reality" (enclose the phrase in braces or quotes to find papers that contain the exact phrase) OR "virtual environment" OR "virtual simulation" OR vr AND evacuation. For Scopus, the search fields were article titles, abstracts, and keywords. For Engineering Village, the search fields were subjects, titles, and abstracts. The searches were not limited to any other constraints, such as language or time span. The searches were conducted on 22 January 2018 and yielded a total of 567 results (including duplicates), 233 of which were from Scopus (www.scopus.com) and 334 from Engineering Village (www.engineeringvillage.com).

After duplicates were removed, a filtering process was carried out following a framework called Preferred Reporting Items for Systematic Literature Reviews and Meta-Analyses (PRISMA) (Moher, Liberati, Tetzlaff, & Altman, 2009). The papers were filtered in accordance with the following inclusion and exclusion criteria: firstly, papers were excluded if there were no IVR SG-based evacuation training and research related to terms in the titles or abstracts. Subsequently, the rest of the papers' full texts were assessed for eligibility. The eligible papers were included if they met all the following criteria:

(i) An IVR SG prototype was proposed, or an existing IVR SG prototype was evaluated and analyzed;

(ii) An experiment was carried out to gather the learning or behavior outcome;

(iii) A data analysis of the outcome was carried out to evaluate and validate the prototype.

The papers were excluded if:

(i) There was no immersive virtual reality principle involved in the prototype because there are other forms of VR (e.g., flat screens showing virtual environment) which do not involve full participant immersion;

(ii) Only the theories, concepts, frameworks, or proposals were discussed without following up experiments or case studies. This research aims to investigate IVR SGs for evacuation from development to implementation. The implementation stage is an important step not only to evaluate and validate IVR SGs, but also to implement them into practice and influence large number of evacuees.

After the filtering process, snowballing was adopted to identify additional papers. The snowballing was based on the previous filtered results after PRISMA. Both backward and forward snowballing was carried out on Google Scholar. The inclusion and exclusion criteria were the same as the ones adopted in the PRISMA framework.

Figure 2 shows diagrammatically the above-mentioned methodological process derived from the PRISMA framework.

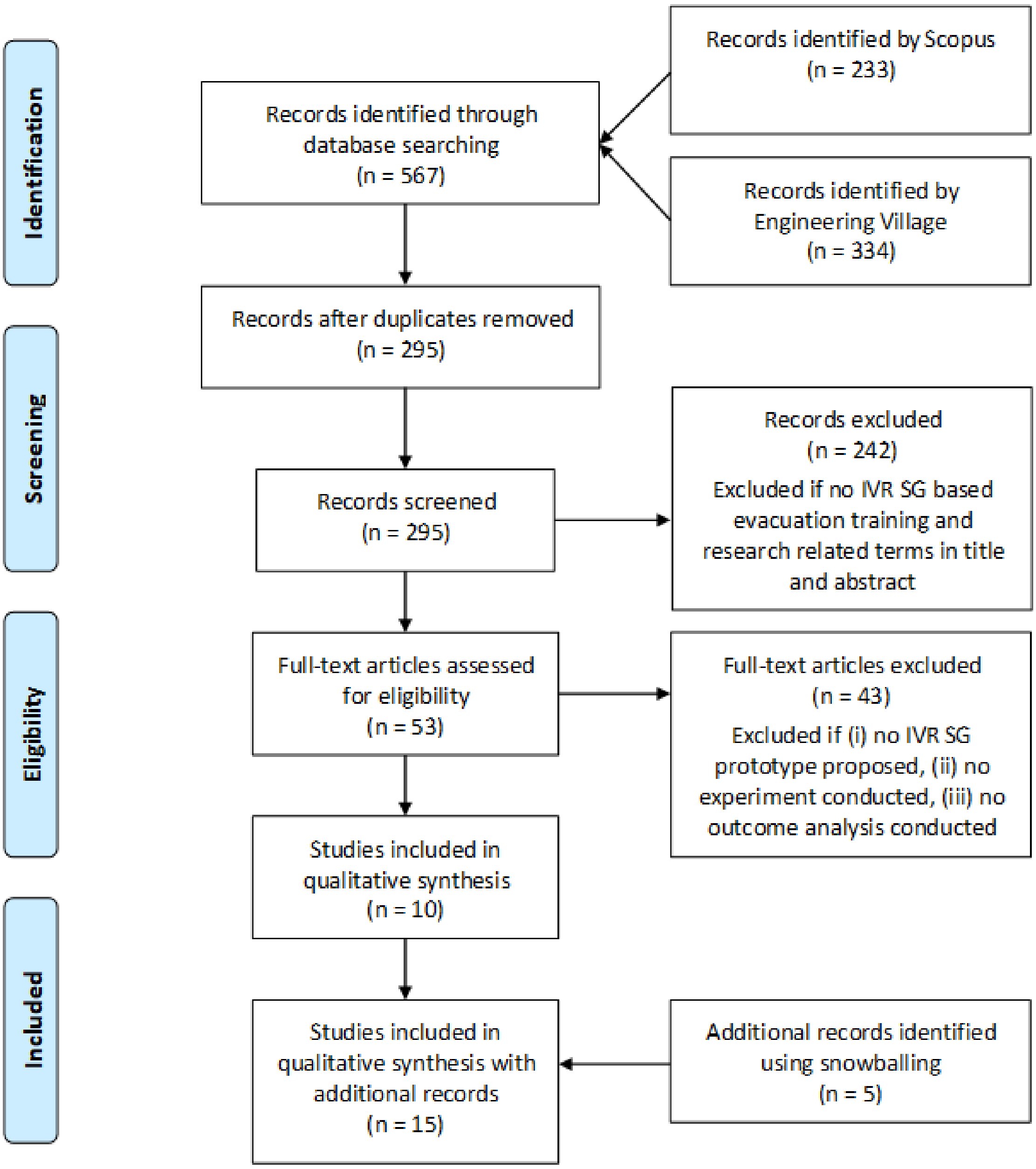

Figure -2 Study selection process

As a result, 15 papers were identified as being relevant to this systematic literature review, which are (Andrée, Nilsson, & Eriksson, 2016; Aoki, Oman, & Natapoff, 2007; Burigat & Chittaro, 2016; Chittaro & Buttussi, 2015; Cosma, Ronchi, & Nilsson, 2016; Duarte, Rebelo, Teles, & Wogalter, 2014; Kinateder, Ronchi, Gromer, Müller, Jost, Nehfischer et al., 2014b; Kinateder, Müller, Jost, Mühlberger, & Pauli, 2014a; Kinateder, Gromer, Gast, Buld, Müller, Jost et al., 2015; Li et al., 2017; Meng & Zhang, 2014; Ronchi, Kinateder, Müller, Jost, Nehfischer, Pauli et al., 2015; Ronchi, Nilsson, Kojić, Eriksson, Lovreglio, Modig et al., 2016; Smith & Ericson, 2009; Zou, Li, & Cao, 2016). Among these papers, 14 were published in journals, with the remaining one published as a conference proceeding. Most of the papers were published after 2014, with only two were published in 2007 and 2009. One possible reason may be that IVR technology has become more popular only in recent years, although the concept of IVR itself can be traced back to the 80's (Fisher, McGreevy, Humphries, & Robinett, 1987). Another interesting finding is that most of the IVR SGs for evacuation training and research have been carried out in Europe, with only three conducted in China and another two in the US.

### 4.3 Assessing the Quality of Studies

After the eligible papers were identified, a scoring process was conducted to assess the quality of the papers. The scoring criteria was derived from the quality assessment approach adopted by Connolly et al. (2012). To be specific, each paper was scored by two authors based on the three assessment questions asked below:

(i) How appropriate is the prototype design for addressing the questions of this review?

(ii) How appropriate are the methods and analysis for addressing the questions of this review?

(iii) How relevant is the focus of the study for addressing the questions of this review?

Each dimension score ranged from 1 to 3 where 1 means low quality, 2 means medium quality and 3 means high quality. Each of the three dimensions' scores was summed to get a total score for each paper. As a result, each paper received two total scores from two raters, and the mean for these two scores was calculated to get the final score of each paper. Possible final scores ranged from 3 to 9, where 3 stands for low quality and 9 stands for high quality.

Each of the 15 papers was coded, and Figure 3 shows a histogram of the final scores. The mean for the 15 papers' ratings is 6.63, and the mode is 6. Twelve papers rated 6 or over were considered as higher quality papers that provided stronger empirical evidence regarding this review's objective. All the papers are summarized in Appendix A with high-quality papers being highlighted, showing the names of the authors, years published, objectives of the study, methods, results and conclusions.

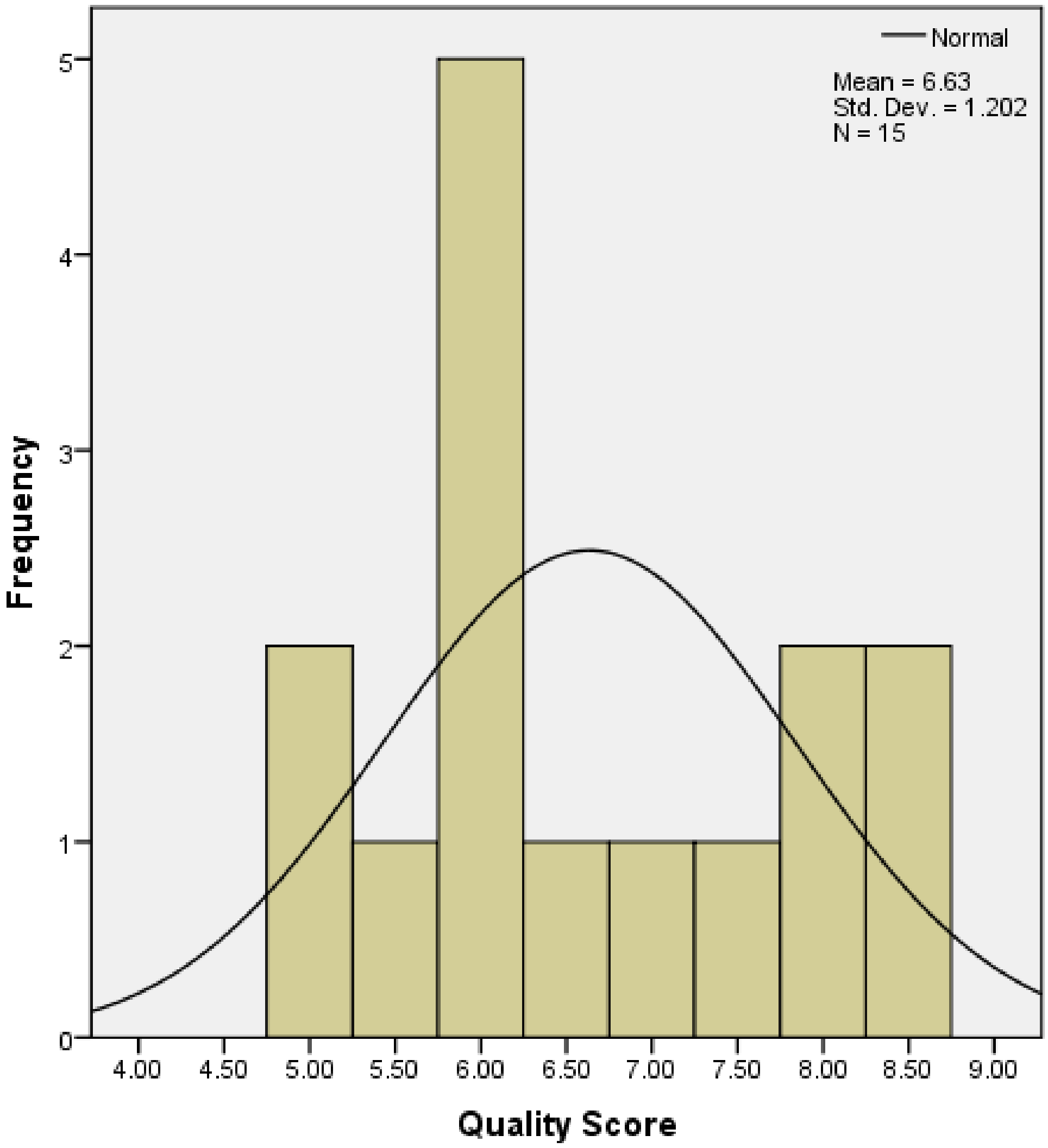

Figure – 3 Quality scores for eligible papers

### 4.4 Summarizing the Evidence

The eligible papers were coded and analyzed using a data extraction spreadsheet that included the research aspects and questions mentioned above. Given that there is only a small number of eligible papers, all the 15 papers are discussed in-depth in this section.

### 4.4.1 Gaming Outcomes and Measures

Two types of outcomes were identified in section 4.1, namely, pedagogical outcomes and behavioral outcomes. Table 2 shows the number of papers that addressed the different outcomes in terms of the simulated emergencies of the games.

Table 2

Gaming outcomes and impacts

| Simulated Events | Pedagogical Outcomes | Behavioral Outcomes | Pedagogical and Behavioral Outcomes |
|---|---|---|---|
| Tunnel Fire | | (Cosma et al., 2016; Kinateder et al., 2014a; Kinateder et al., 2014b; Kinateder et al., 2015; Ronchi et al., 2015; Ronchi et al., 2016) | |
| Building Fire | (Smith & Ericson, 2009) | (Andrée et al., 2016; Duarte et al., 2014; Meng & Zhang, 2014) | |
| Aviation Emergency | (Burigat & Chittaro, 2016; Chittaro & Buttussi, 2015) | | |
| Spacecraft Emergency | (Aoki et al., 2007) | | |
| Building Earthquake | (Li et al., 2017) | (Li et al., 2017) | (Li et al., 2017) |

Among the eligible papers, there were four papers that were identified to use IVR SGs as a pedagogical tool, while nine papers were identified to use IVR SGs as a behavioral tool. One paper used IVR SGs as both a pedagogical and behavioral tool.

#### 4.4.1.1 Pedagogical Impact

In five studies, IVR SGs were implemented as pedagogical tools (Aoki et al., 2007; Burigat & Chittaro, 2016; Chittaro & Buttussi, 2015; Li et al., 2017; Smith & Ericson, 2009).

In the reviewed papers, various evacuation knowledge was delivered as learning outcomes. In total, three types of knowledge were identified by this review, namely evacuation best practices (Chittaro & Buttussi, 2015; Smith & Ericson, 2009), self-protection skills (Li et al., 2017), and spatial knowledge (Aoki et al., 2007; Burigat & Chittaro, 2016).

Smith and Ericson (2009) encouraged participants to identify potential fire hazards. Following that, participants were educated on the best practices for evacuation in case of fire at home. A significant increase in measured fire-safety knowledge after training was observed in the results. Chittaro and Buttussi (2015) provided the entire evacuation protocol for an aviation emergency ranging from turbulence instability response to evacuation into a life raft after a forced landing at sea. As a result, participants were trained with a total of ten learning items associated with the best practices for an aviation emergency. Results showed that IVR SG had better performance in terms of knowledge retention after one week compared to the safety card method. This is critical for people to recall knowledge on proper evacuation behaviors in real emergencies.

Apart from the studies focusing on multiple knowledge (e.g., best practices), three other papers focused on single knowledge transfer. Li et al. (2017) proposed using IVR SG as a training tool to teach participants how to protect themselves in common indoor environments during an earthquake. On average, participants trained by IVR SG performed better than those trained by safety videos or manuals in terms of hazard awareness and avoidance. Burigat and Chittaro's (2016) main learning goal was to let participants acquire spatial knowledge to permit effective evacuation of an airplane. Results showed that participants trained by IVR SG obtained better spatial knowledge than those trained by safety cards. Aoki et al. (2007) investigated spatial skills influenced by relative body orientation during IVR training in a spacecraft evacuation. Results showed that local training (visually upright relative to the "local" module) enabled landmark and route learning, while station training (constant orientation irrespective of local visual vertical) improved sense of direction and performance in low visibility.

The reviewed papers show that IVR SGs have the capability of delivering certain amount of evacuation knowledge as learning outcomes, even as complex as best practices with multiple learning items. The expected learning outcomes, which are the basis for game storyline and narrative, should be defined in prior to IVR SGs development.

After the IVR SGs training session, the learning performance of all participants mentioned previously (Aoki et al., 2007; Burigat & Chittaro, 2016; Chittaro & Buttussi, 2015; Li et al., 2017; Smith & Ericson, 2009) was assessed bearing in mind the IVR SGs learning outcomes. Several assessment measures were recognized, and these are shown in Table 3.

Table 3

Learning measures of IVR SGs

| Learning Measures | Knowledge Acquisition | Knowledge Retention | Both Knowledge Acquisition and Retention |
|---|---|---|---|
| Questionnaire | (Smith & Ericson, 2009) | | |
| Open-ended question interview | (Aoki et al., 2007; Chittaro & Buttussi, 2015) | (Chittaro & Buttussi, 2015) | (Chittaro & Buttussi, 2015) |
| Paper-based test | (Burigat & Chittaro, 2016) | | |
| Logged game data (e.g., evacuation time, damages received) | (Aoki et al., 2007; Burigat & Chittaro, 2016; Li et al., 2017) | (Li et al., 2017) | (Li et al., 2017) |

To assess participants' knowledge acquisition, Smith and Ericson (2009) measured how many correct answers were given to the questions related to evacuation safety knowledge before and immediately after the IVR SGs training. Aoki et al. (2007) asked participants to verbally describe the configuration of the spacecraft after the training, thus, to assess their spatial knowledge. Chittaro and Buttussi (2015) asked participants to orally answer the open-ended questions related to evacuation safety knowledge before and immediately after the IVR SGs training to avoid suggesting possible answers such as a multiple-choice questionnaire would do. In another study, Burigat and Chittaro (2016) adopted a paper-based approach using maps to let participants mark positions of exits and their initial seats as they aimed to gain spatial knowledge. In addition to the paper-based approach, Burigat and Chittaro (2016) also recorded the evacuation time of participants to evaluate their learning results. Aoki et al. (2007) recorded evacuation time, numbers of turns and errors to assess participants' spatial knowledge acquisition. Li et al. (2017) used logged game data in terms of the physical damage received during the training session to evaluate participants' learning performance, thus, to provide feedback in order to improve self-protection skills. In accordance with learning outcomes, different assessment measures can be applied.

Smith and Ericson (2009) argued that there was a significant improvement in measured fire-safety knowledge after training. Chittaro and Buttussi (2015) revealed that IVR SGs are superior to the traditional approaches on knowledge retention. This is the fundamental requirement for survival during evacuation because people need to retain correct evacuation procedures over a long-time span so that to apply them whenever facing an emergency. Li et al. (2017) and Burigat and Chittaro (2016) found IVR SGs could produce better knowledge transfer than traditional approaches. Therefore, it can be concluded that IVR SGs are effective in delivering considerable evacuation knowledge, no matter whether it is multiple knowledge (e.g., best practices) or single knowledge (e.g., spatial skill).

After running the entire IVR SGs training session, researchers from two studies (Chittaro & Buttussi, 2015; Li et al., 2017) conducted second-round tests one week later to assess participants' knowledge retention. With the learning measures adopted for assessing knowledge acquisition in the first tests, the same measures were applied again in the second tests to observe the changes in evacuation knowledge between the first and second tests so that the knowledge retention can be evaluated. Both studies showed that in the context of evacuation training, participants that received training by IVR SGs had better performance in terms of knowledge retention compared to those that were trained using traditional approaches.

#### 4.4.1.2 Behavioral Impact

In ten studies, the IVR SGs approach was adopted as a behavioral analysis tool (Andrée et al., 2016; Cosma et al., 2016; Duarte et al., 2014; Kinateder et al., 2014a; Kinateder et al., 2014b; Kinateder et al., 2015; Li et al., 2017; Meng & Zhang, 2014; Ronchi et al., 2015; Ronchi et al., 2016). In terms of the behavioral outcomes and measures, they are rather diverse due to the various purposes of behavioral analysis. Table 4 shows the different purposes recognized by this review.

Table 4

Behavioral outcomes of IVR SGs

| Behavioral Outcomes | | Description |
|---|---|---|
| Evacuation facility validation | (Andrée et al., 2016; Cosma et al., 2016; Ronchi et al., 2016) | Test and validate different evacuation facility designs and installations |
| Behavioral compliance | (Duarte et al., 2014) | Investigate whether participants follow the evacuation instructions |
| Hazard awareness | (Li et al., 2017) | Investigate whether participants can notice hazards in the environment |
| Behavior validation | (Ronchi et al., 2015) | Validate hypothetical behavior model |
| Social influence | (Kinateder et al., 2014b; Kinateder et al., 2014a) | Investigate social influence on evacuation behavior |
| Behavior recognition | (Kinateder et al., 2015) | Recognize different behavior under different evacuation conditions |
| Way-finding behavior | (Meng & Zhang, 2014) | Explore evacuation way-finding behavior |

Cosma et al. (2016) investigated the impact of evacuation lighting systems for rail tunnel evacuation. Participants were exposed to an emergency evacuation scenario, and their movement paths and evacuation time were recorded to evaluate different way-finding lighting installations impacts. Results showed that both dynamic alternate and continuous way-finding lighting systems had a positive impact on evacuation, and no significant differences were found between these two systems. Ronchi et al. (2016) put participants in IVR SGs to test different designs of flashing lights at emergency exit portals. Then participants were asked to finish a questionnaire to provide recommendations on different portal designs. Various variables were investigated.

Results suggested that participants preferred green and white flashing lights rather than blue lights, at a flashing rate of 1 and 4 Hz rather than 0.25 Hz, a light emitting diode light source rather than single and double strobe lights. No significant preference difference was found between different numbers and layouts of lighting on portals. Andree et al. (2016) proposed using IVR SGs to validate the evacuation procedures for elevators in high-rise buildings by analyzing participants' exit choices and waiting times. Results suggested that participants would more likely evacuate by elevators influenced by the green flashing way-finding lighting system. Moreover, participants tended to wait for elevators either a limited time (<5 min) or a long time (>20min). Duarte et al. (2014) studied participants' behavioral compliance by counting the number of times participants followed the directions indicated by the evacuation signs. Results showed that evacuation signs were effective in changing evacuees' behavior. Li et al. (2017) investigated whether participants noticed hazards around them by tracking their visual attention. The percentage of dangerous objects noticed by participants during the IVR SGs was calculated to assess participant behavior in response to an emergency. Results indicated that participants trained by IVR SG before were able to notice more hazards than those trained by other approaches such as videos and manuals. Ronchi et al. (2015) proposed an IVR SG to perform evacuation model validation by comparing participants' actual movement paths to the hypothetical paths. Results showed that hypothetical paths based on the shortest distance employed by evacuation models might be over-simplified compared to actual movement paths. Kinateder et al. (2014b; 2014a) used IVR SGs to investigate social influences on evacuation behavior. The non-playable characters (NPCs) inside the IVR SGs undertook social interactions with participants. Participants' behavior was analyzed by evaluating destination choice, movement pathway, and evacuation time to explore social influences on their decision-making processes. Results indicated that other evacuees had a strong influence on participants evacuation behavior. Participants were more likely to react similarly to the NPCs. Kinateder et al. (2015) let participants face burning dangerous goods vehicle or burning heavy goods vehicle to explore different outcomes of evacuation behavior under different conditions. Results showed that participants had similar patterns of evacuation behavior under both conditions, while they still perceived significantly more danger in the burning dangerous goods condition. Meng and Zhang (2014) argued that IVR SGs were capable of exploring the evacuation way-finding behavior by analyzing participants' evacuation time, route choice, and movement pathway. Results suggested that participants had poor way-finding performance in a fire emergency, which may be resulted from high physiological and psychological stress.

Benefit from the nature of IVR SGs, absorbing gaming environment is able to induce participants to react as close as to the reactions in real life. The empirical evidence from the reviewed papers demonstrates that IVR SGs are a promising tool to analyze participants' behavior for different purposes. As for the behavior measures, the choose of different approaches is heavily relying on different expected behavior outcomes. Considering that, IVR SGs need to integrate the corresponding functions to carry out the analysis for pre-defined behavior outcomes.

#### 4.4.1.3 Participation Experience

A few studies investigated participation experience. It can be categorized into two different aspects, which are self-reported psychological assessment and device-based physiological assessment. In terms of the psychological aspect, Chittaro and Buttussi (2015), Smith and Ericson (2009), Burigat and Chittaro (2016), and Meng and Zhang (2014) introduced questionnaires to assess self-reported fear, engagement, stress, usability and workload. Zou et al. (2016) applied the positive affect and negative affect scale (PANAS) to let participants describe emotions. In relation to the physiological aspect, Chittaro and Buttussi (2015) provided an electrodermal activity sensor (EDA) to track skin conductance levels (SCL) to evaluate fear and anxiety (an increase in SCL indicates arousal) , and a photoplethysmograph sensor (PPG) to obtain blood volume pulse amplitude (BVPA), which can be employed as an index of sympathetic arousal (a decrease in BVPA indicates increased arousal). Meng and Zhang (2014) tracked participants skin conductivity and heart rate in order to give an insight about the stress of participants when they were undertaking an IVR SGs session. Zou et al. (2016) applied a multichannel physiological recorder to track participants skin conductivity in order to measure emotional responses of participants when facing a fire emergency.

Participation experience can be measured to evaluate IVR SGs. Self-reported psychological assessment can be applied to investigate the usability of IVR SGs while device-based physiological assessment can be applied to reveal the subconscious activities as strong feasibility evidence to support IVR SGs.

### 4.4.2 Gaming Environment

Six major components for the gaming environment were raised in the previous questions, namely teaching method, navigation solution, narrative method, hazard simulation, senses stimulation, and non-playable characters (NPCs).

#### 4.4.2.1 Teaching Methods

Two types of teaching methods were used to deliver evacuation knowledge. One was to give feedback after the response to an event (Burigat & Chittaro, 2016; Chittaro & Buttussi, 2015; Li et al., 2017). In this way, participants could know if their behavior was right or wrong after making a decision to an event, thus, participants could learn from their mistakes. Another one was to provide instructions before the response to an event (Aoki et al., 2007; Smith & Ericson, 2009). In other words, participants were told what to do before dealing with an event, thus, to learn the appropriate behavior accordingly.

Regarding feedback after the response, there are two forms of feedback identified. One is immediate feedback. To be specific, if participants make an inappropriate action, they will be informed by the written or verbal messages immediately while in the game (Burigat & Chittaro, 2016; Chittaro & Buttussi, 2015). In this way, participants can rectify their actions and memorize them during the IVR SGs simulation. The second form of feedback is post-game feedback, which means participants only receive feedback after the completion of IVR SGs simulation. For instance, participants can evaluate their evacuation responses based on the final results they received after the training (Li et al., 2017). Irrespective of how it is provided, feedback is fundamental to enhancing the knowledge acquisition of participants because it will allow them to learn from their mistakes and rectify their responses to actual emergency situations.

Another teaching method was identified by this review, which is to provide instructions ahead to participants. Smith and Ericson (2009) applied instructions to assist participants in advance of what to do facing following up emergency situations. By doing so, it may reduce fear and stress, hence, help participants focus on training materials and complete training process since Smith and Ericson (2009) found that their participants (7 – 11 years old children) were nervous and fearful when facing fire emergencies without telling them what to do in their previous study. Aoki et al. (2007) gave instructions to participants as training tours in order to help them get familiar with evacuation paths since their research was focusing on teaching spatial knowledge.

No matter which teaching method was applied, each study confirmed a positive learning outcome regarding evacuation knowledge acquisition. However, the difference of the efficiency between each method is still lacking in the literature. In the reviewed papers, both teaching methods were applied only in accordance with participants background.

#### 4.4.2.2 Navigation Solutions

The navigation solution is a critical factor related to a side effect of IVR SGs commonly known as motion sickness. Participants suffer from sensory conflicts when the perceived movement inside IVR SGs is not corresponded with their physical bodies' motion (Hettinger & Riccio, 1992). In this review, only one study was identified employing a questionnaire to evaluate motion sickness of participants (Smith & Ericson, 2009). It was shown that 20% participants found motion sickness to be a disagreeable side effect during the game. However, the detailed navigation solution was not stated in the study. This study only mentioned that participants could navigate by manipulating a gamepad. Other four studies provided detailed navigation solutions (Andrée et al., 2016; Aoki et al., 2007; Burigat & Chittaro, 2016; Chittaro & Buttussi, 2015; Duarte et al., 2014). There were two different solutions identified by this review. One was to move forward/backward by tilting a joystick forward/backward, and rotate to left and right by tilting a joystick left and right (Aoki et al., 2007; Burigat & Chittaro, 2016; Duarte et al., 2014). The other was only to move towards the facing direction by holding a button on a joystick (Andrée et al., 2016; Chittaro & Buttussi, 2015). By doing this, participants need to turn their heads together with their bodies every time they wanted to turn to a new direction in the game.

The navigation solution is related to the IVR hardware. Different hardware has different representation and control system. Even for the same hardware, there are still different navigation solutions available by manipulating joysticks differently as we identified in the preceding paragraph. However, at this stage, there is no study investigating the impact of different navigation solutions on motion sickness within the scope of this research.

#### 4.4.2.3 Narrative Methods

The narrative method refers to the method applied to encourage participants to follow and complete storylines of IVR SGs. The storyline is a set of scenarios in which participants can make decisions and take actions accordingly. In this way, the expected outcomes can be generated and achieved. The underlying reason for a specific narrative is that an IVR environment can often be simulated as an open world where participants may get lost or wander around out of curiosity (Lovreglio et al., 2018). As a result, there were four narrative methods recognized by this review, namely action-driven method (Chittaro & Buttussi, 2015), instruction-driven method (Andrée et al., 2016; Aoki et al., 2007; Burigat & Chittaro, 2016; Cosma et al., 2016; Meng & Zhang, 2014; Smith & Ericson, 2009; Zou et al., 2016), performance-driven method (Li et al., 2017), and surrounding-driven method (Duarte et al., 2014; Kinateder et al., 2014a; Kinateder et al., 2014b; Kinateder et al., 2015; Ronchi et al., 2015; Ronchi et al., 2016).

The action-driven method means that storylines can be driven by a sequence of actions taken by participants. Participants have to select the correct actions from very limited movement choices in order to make progress through the game (Chittaro & Buttussi, 2015). The second solution is called instruction-driven method. Prior instructions are provided to guide participants' behavior while they can still move freely in the IVR environment (Andrée et al., 2016; Aoki et al., 2007; Burigat & Chittaro, 2016; Cosma et al., 2016; Meng & Zhang, 2014; Smith & Ericson, 2009; Zou et al., 2016). Additionally, Burigat and Chittaro (2016) displayed instructions to bring participants back on right track if they stopped or moved toward a wrong direction. Apart from the two methods discussed above, the performance-driven method was also identified by this review. Li et al. (2017) introduced a scoring system, which can generate scores based on participants' gaming performance. By doing this, participants were encouraged to complete the storyline and do their best during the training sessions. The last method identified is called surroundings-driven method. To be specific, the IVR environment is filled with simulated hazards, leaving limited possible directions participants can move in. Eventually, the entire IVR environment will be filled with simulated hazards. Therefore, the possible movement area is so restricted that participants have no choice other than to follow and complete the pathway determined by the storyline (Duarte et al., 2014; Kinateder et al., 2014a; Kinateder et al., 2014b; Kinateder et al., 2015; Ronchi et al., 2015; Ronchi et al., 2016).

The narrative methods can largely influence on how the game is progressed by participants. Both action-driven and surroundings-driven methods make the IVR environment limited to move around leaving participants no other choices but have to follow the storyline. In opposite, instruction-driven and performance-driven methods still keep the open world of IVR SGs. The progress of the storyline is relying on participants personal ability and willingness to complete the game.

#### 4.4.2.4 Hazards Simulation

In the reviewed studies, several types of hazards were adopted and simulated. These hazards can be categorized to two different levels: static level and dynamic level.

The static level means that the simulated hazards are static, and they do not interact with participants; hence, they do not negatively affect participants. These hazards can be used to increase fear perception, trigger events, constrain moving area, and improve environmental realism. For instance, a burning vehicle that blocks one evacuation way (Kinateder et al., 2014b; Kinateder et al., 2015; Ronchi et al., 2015), an engine failure (Chittaro & Buttussi, 2015) or an explosion (Duarte et al., 2014) triggers the evacuation event, and high-fidelity fire simulation promotes a level of realism (Zou et al., 2016).

The dynamic level is a more advanced representation of the simulated hazards. They can interact with participants and impact participants by providing negative experience. For instance, participants can see their eyes spattered by blood if they have been hit on the head, or they can believe they are in danger of drowning if water enters the cabin of a plane (Chittaro & Buttussi, 2015). If participants are standing in an environment with smoke, they will lose visibility (Aoki et al., 2007; Burigat & Chittaro, 2016; Chittaro & Buttussi, 2015; Cosma et al., 2016; Duarte et al., 2014;

Kinateder et al., 2014a; Kinateder et al., 2014b; Kinateder et al., 2015; Ronchi et al., 2015; Smith & Ericson, 2009). In this case, Smith and Ericson (2009) suggested participants crawl on the floor to get under the smoke as the correct behavior and response. In another study (Li et al., 2017), falling and fragile objects can hurt participants' virtual bodies during an earthquake simulation. Therefore, participants need to come up with a strategy to protect themselves. No matter what type and level of hazards are chosen, participants are exposed to these realistic hazards in a completely safe environment. In this way, IVR SGs can largely influence participants' behavior.

One of the advantages of IVR SGs is that various hazards and dangerous situations, which are impossible to be presented in real life, can be presented putting no one's life at risk. On the one hand, a large number of static hazards can be applied to make progress of the game and make the IVR environment realistic. On the other hand, dynamic hazards can be adopted to create negative experience such as hurt, bleeding, or drowning in order to enhance participants' impression regarding the consequences of inappropriate responses during an emergency.

#### 4.4.2.5 Senses Stimulation

Another major component of the gaming environment is the senses stimulation. Four different types of stimulation were identified by this review. Obviously, every study had both visual and auditory stimulations since IVR SGs are basically video games. Apart from that, motion interaction (Li et al., 2017; Smith & Ericson, 2009) and olfactory stimulation (Meng & Zhang, 2014) were also recognized by this review. Li et al. (2017) introduced the motion interaction system to allow participants to physically undertake the following actions: moving, crouching, and head-protecting during the IVR SGs session. Smith and Ericsson (2009) enabled participants to crawl physically in order to get under the smoke in the IVR SGs session. By using motion tracking systems, participants can obtain a better engagement and perception in comparison to the use of traditional joysticks or joypads. In addition, it is more user-friendly as it is easier to control and understand. Another interesting application is that Meng and Zhang (2014) proposed a smoke generator to provide olfactory and visual stimuli to participants with real smoke in compliance with the IVR SGs session. In this case, participants received a relatively high-fidelity IVR experience. The combination of different sense stimulations is beneficial to provide a better engagement in order to minimize unrealistic behavior in the virtual environment.

#### 4.4.2.6 Non-playable Characters

Two types of NPCs were identified by this review, non-interactive NPCs and interactive

NPCs. Chittaro and Buttussi (2015) introduced non-interactive NPCs only to represent other evacuees in order to improve the realism of the IVR experience. Interactive NPCs were also proposed in the same study to provide recommendations to help participants in order to achieve the intended learning outcomes. Kinateder et al. (2014a; 2014b) adopted interactive NPCs to perform social interaction with participants to investigate social influences on evacuation behavior. To be specific, these NPCs performed various actions such as running to a wrong evacuation direction so that participants' evacuation behavior can be influenced.

Representing NPCs is a challenge for the development of IVR SGs (Lovreglio et al., 2017). The reviewed papers show that IVR SGs have the possibility to investigate how participants' behavior is influenced by other evacuees. The use of NPCs can help deliver expected outcomes.

## 4.5 Interpreting the Findings

Based on the data analyzed above, we summarized the various factors and aspects influencing the development and implementation of an IVR SGs evacuation research into a conceptual framework, shown in Figure 4.

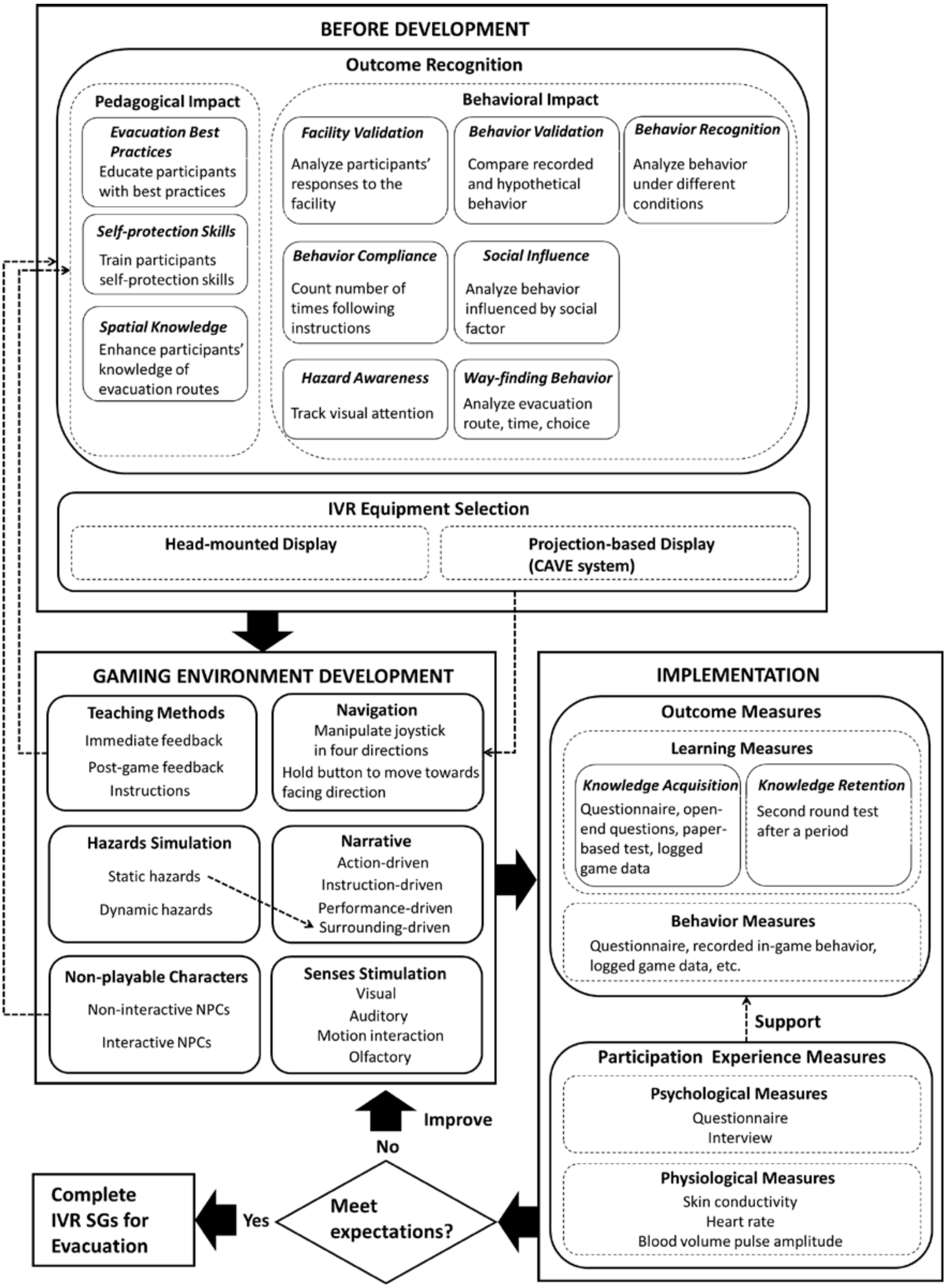

Figure 4 – A conceptual framework for developing and implementing IVR SGs for evacuation research

Before development process, researchers need to identify expected outcomes and impacts to be achieved by IVR SGs. In order to achieve the goals of pedagogical impact, pre-defined evacuation knowledge needs to be delivered to participants effectively.

There are two teaching methods that are suitable for different target groups, which are feedback after response and instruction before responses. Both methods are proven effective. The selection criteria can be based on participants' background and game narrative method. After training, multiple learning measures can be implemented to evaluate learning outcomes. The IVR SGs development also needs to include data recording function if such data (e.g., movement path, evacuation time, exit choice) is necessary for learning measures. For the behavioral impact, various outcomes can be acquired by the IVR SGs approach such as behavior compliance, behvaior recognition, behavior validation. In accordance with different behavioral outcomes, appropriate measures need to be considered in the gaming development process in case further analysis tools are required to be added into IVR SGs in order to track behavior and record data.

Apart from the outcomes to be identified before the development, another important consideration is the IVR equipment selection. Two types of equipment were identified in this review, namely head-mounted display (HMD) and projection-based display (PBD). Each equipment type requires different navigation solutions, which are the essential parts of IVR SGs. Appropriate navigation solution should be designed thoroughly in accordance with the available IVR equipment in order to deliver expected outcomes by providing a comfortable gaming experience (Riecke, Bodenheimer, McNamara, Williams, Peng, Feuereissen, 2010).

During the development process, a few elements in the framework can be taken into consideration by developers in order to achieve high-quality IVR SGs. Simulated hazards are mainly related to simulation events. Static hazards can influence game narrative, especially the surrounding-driven method. Static hazards can make up dangerous area leave participants limited moving options so that they have to follow the pre-defined game storylines. Dynamic hazards can be rather flexible depending on the research purposes and participants' backgrounds. Narrative methods can significantly influence game storyline development. Participants' background, expected outcomes, and IVR environment should be taken into account when deciding the appropriate narrative method. Senses stimulation is mainly based on available equipment. It is believed that participants should feel being physically inside the virtual world in order to minimize unrealistic behavior (Rüppel & Schatz, 2011). The use of multiple senses stimulation can promote this influence. NPCs are also an important element of IVR SGs. Not only they can help increase the realism of IVR SGs, but also they can interact with participants to help achieve certain research goals.

The implementation phase of IVR SGs follows the development phase. During implementation, expected outcomes can be measured to validate the effectiveness of IVR SGs. Along with the outcomes measuring, participation experience can also be obtained as strong evidence to support the outcomes, and to provide valuable feedback to assess the usability of IVR SGs. If the outcomes or the experiences are not satisfactory, IVR SGs need to be improved in accordance with the measured results until the results meet the expectations. After that, a final product of IVR SGs is completed which is able to deliver the expected outcomes meeting research requirements.

## 5. Conclusions

We carried out a systematic literature review on IVR SGs for evacuation training and research. The pedagogical and behavioral outcomes, gaming environment development, and outcomes and participation experience measures were extensively explored. The findings indicate the advantages and disadvantages of IVR SGs in terms of delivering evacuation knowledge and conducting evacuation behavior analysis. This study provides insights into the characteristics and structure of IVR SGs. As a result, we proposed a conceptual framework for developing and implementing IVR SGs based on the investigation into the existing literature. This framework aims to contribute to future applications of IVR SGs for evacuation training and research.

This review has a few limitations. The review excluded theoretical papers without following up case studies. As a result, all the analyzed data involves empirical evidence extracted from the prototypes given in the existing literature. There may be other aspects that are important to developing and implementing successful IVR SGs for evacuation training and research. Furthermore, the conceptual framework is derived from systematic literature review findings. Regarding that, this framework must be further studied and validated.

When conducting this study, we found some potential directions for future research. As we stated before, there is still in need to investigate the impact of different navigation solutions on motion sickness. Apart from that, within the 15 reviewed papers, eleven focused on fire evacuation, three focused on aviation and spacecraft evacuation, while only one refers to evacuation during earthquakes. There seems to be a significant gap between the research on fire evacuation and research on earthquake evacuation, with little attention given to the latter. Regarding that, more attention needs to be paid on earthquake safety training (Lovreglio et al., 2017). Another interesting finding is that only one study took children as research subjects (Smith & Ericson, 2009). The rest of the researches were carried out in universities, with students and staff comprising the majority of the subjects. During an emergency, children are more vulnerable than adults. Therefore, it would be valuable to conduct more research on children using an IVR SG approach.

## Appendix A. Summary of the eligible papers (higher quality papers marked with *).

| Authors and Years | Objectives of study | Methods | Results and Conclusions |
|---|---|---|---|
| (Aoki et al., 2007) | To examine the influence of relative body orientation and individual spatial skills during VR training on a simulated emergency egress task. | 36 subjects were each led on 12 tours through a space station by a virtual tour guide. Subjects wore a head-mounted display and controlled their motion with a game-pad. Groups were balanced on the basis of mental rotation and perspective-taking test scores. Subjects then performed 24 emergency egress testing trials without the tour guide. Smoke reduced visibility during the last 12 trials. Egress time, sense of direction (by pointing to origin and destination) and configuration knowledge were measured. | Both individual 3D spatial abilities and orientation during training influence emergency egress performance, pointing, and configuration knowledge. Local training facilitates landmark and route learning, but station training enhances sense of direction relative to station, and, therefore, performance in low visibility. We recommend a sequence of local, followed by station, and then randomized orientation training, preferably customized to a trainee's 3D spatial ability. |
| (Smith & Ericson, 2009)* | To make fire safety fun and engaging to learn by using game-based VR techniques, while actually helping children remember the steps they need to take to save themselves and others if they ever find themselves in a real fire emergency. | The simulation was run in a four-sided CAVE. Participants were drawn from area Boy Scout troops. Participants were divided into two groups. The control group, which consisted of 10 children, took a post-quiz prior to using the VR application, while the experimental group, which consisted of 12 children, used the VR application first and then took the post-quiz. | From project results, the investigators demonstrated that immersive VR systems have a built-in advantage over prior fire-safety training methods for children. They allow children to experience realistic virtual ''hands-on'' and ''on-site'' experiences for high-risk safety training, which cannot be achieved through lectures or regular video games, e.g., how to kneel down or crawl in a fire situation. In addition, study results show that game- |
| | | | based VR systems increase children's motivation over more traditional teacher–learner forms of VR-based instruction, which was the primary goal of the study. |
| (Duarte et al., 2014)* | To examine how dynamic features in signage affect behavioral compliance during a work-related task and an emergency egress. | All tasks were performed in an immersive virtual reality system - ErgoVR. Ninety participants performed a work-related task followed by an emergency egress. | Although dynamic presentation produced the highest compliance, the difference between dynamic and static presentation was only statistically significant for uncued signs. Uncued signs, both static and dynamic, were effective in changing behavior compared to no/minimal signs. Findings are explained based on sign salience and on task differences. If signs must capture attention while individuals are attending to other tasks, salient (e.g., dynamic) signs are useful in benefiting compliance. This study demonstrates the potential for IVEs to serve as a useful tool in behavioral compliance research. |
| (Kinateder et al., 2014a) * | To examine if and how conflicting social information may affect evacuation in terms of delayed and/or inadequate evacuation decisions and behaviors. | VR scenes were presented stereoscopically by two video projectors. Participants wore passive circularly polarized glasses for 3D effects. Forty participants were repeatedly situated in a virtual reality smoke filled tunnel with an emergency exit visible to one side of the participants. Four social influence conditions were realized. In the control condition participants were alone in the tunnel, while in the other three experimental conditions a virtual agent (VA) was present. | In conclusion, this study demonstrated clear social influence effects on evacuation behavior during tunnel fires, and that social influence may have both positive and negative effects. Increase or decrease of ambiguity about the optimal escape route due to social influence can increase or decrease the frequency of wrong decisions and/or of pre-movement times, respectively, and, hence, can modulate the likelihood of adequate safety behavior. We observed that only about 60% of the participants reacted adequately in a tunnel fire situation and moved toward the next emergency exit independent of social influence. Furthermore, it is alarming to note that even if emergency exits are clearly visible and other people also move there, some individuals still decide to move into the smoke. |
| (Kinateder et al., 2014b) * | To examine the social influence on route choice in a virtual reality tunnel fire. | Two experimental groups were immersed into a virtual road tunnel fire. In the SI group participants saw a virtual agent moving on the shortest route to the nearest emergency exit. In the control group, participants were alone. Destination and exit choices were analyzed using functional analysis and inferential statistics. | SI affected route choice during evacuation but not destination choice: There were no group differences regarding destination choice. Participants in the SI group were more likely to choose a route similar to the virtual agent. Participants in the control group were more likely to choose a longer route along the tunnel walls. |
| (Meng & Zhang, 2014) * | To study people's way-finding behaviour and response during a fire emergency in a virtual environment. | This study used a panorama manifestation system (PM system) to provide the VE. The PM system consists of six 47-inch LCD monitors, a computer with a data collection system and a revolving chair with controlling devices. Forty participants (20 male and 20 female), aged between 20 and 25 (mean = 22.5, SD = 1.8), were recruited through online posters. They were divided into two groups, the control group and the treatment group. | People in fire emergency experienced higher physiological and psychological stress, had different perception style, accordingly had different behaviour patterns and poorer way-finding performance, as compared with people in normal condition. |
| (Chittaro & Buttussi, 2015) * | To assess learning immediately after the experience and knowledge retention over a longer time span. | 48 participants (26 M, 22F) were assigned to the two conditions in such a way that:(i) the proportion of men and women was identical (13M and 11F in each group), since gender-balanced groups are particularly important in studies where fear is involved, (ii) the two groups were very similar in terms of age, frequency of air travel, frequency of video game use, and fear of flying. | The paper obtained results for each of the three topics: (i) to the best of our knowledge, we proposed the first immersive serious game for aviation safety education of the general public, (ii) the experimental evaluation showed that, unlike the safety card, the immersive serious game produces in users a knowledge gain that is maintained after one week (people who used the card suffered instead a statistically significant loss of the acquired knowledge after one week), (iii) the immersive game was able to produce more engagement, negative emotion (fear) and physiological arousal than the safety card, a factor that can contribute to explain its positive impact on knowledge retention. |
| (Ronchi et al., 2015) * | To increase the understanding on evacuation behaviour by a case study on the analysis of evacuation travel paths in virtual reality (VR) tunnel fire experiments. | The study was conducted in the 3D-multisensory CAVE laboratory. Twenty-one participants were performing the experiment. | The proposed method is a useful tool to study human behaviour during tunnel evacuations and the subsequent development and validation of the assumptions adopted by computational modelling tools (i.e., evacuation models). The development of robust models permits an increased accuracy in risk assessment in the case of tunnel fire evacuations and the definition of optimal design solutions for tunnel safety. |
| (Kinateder et al., 2015) * | To investigate the effect of an increased risk during a simulated tunnel emergency on participants' subjective hazard perception and evacuation behavior. | Using a five sided CAVE system, two experimental groups were immersed into a virtual road tunnel fire emergency. In the dangerous goods condition a burning gasoline transporter was visible. In the control condition a burning heavy goods vehicle was visible. Hazard perception, pre-movement time, movement time and exit choices were analyzed. | Results: In the dangerous goods condition the situation was rated significantly more dangerous than in the control condition. In both conditions participants showed appropriate behavioral reactions and either moved to an emergency exit or to an emergency phone. |
| (Andrée et al., 2016) * | To study exit choice and the waiting time for evacuation elevators in high rise buildings. | A Cave Automatic Virtual Environment (CAVE) system was used to display the environment. A total of 72 participants took part in the study, 29 women and 43 men. Their ages ranged from 18 to 69 years, and the average age of the participants was 26.5 years. | Results suggest that a simple way-finding system using green flashing lights can influence people to more likely choose the elevator as their first evacuation choice. The results also show that the general trend is that people wait for either a limited time (<5 min) or a long time (>20 min). |
| (Burigat & Chittaro, 2016) * | To leverage the power of VEs to create novel tools for emergency evacuation preparedness and to evaluate on users if such tools are actually more effective than the printed maps currently in use. | We carried out a lab study comparing a version of the tool with Active navigation, a version with Passive navigation, and a traditional Safety Card. We recruited a total of 54 participants (44 male, 10 female) among undergraduate Computer Science students at our university. | Results of our study show that the VE-based approach produces objectively better spatial knowledge when users are asked to pinpoint their assigned position in the environment, and that active navigation produces a performance improvement in a subsequent virtual evacuation. Moreover, the VE-based approach is perceived as more enjoyable, easier to comprehend and more effective than printed maps when active navigation is available. |
| (Ronchi et al., 2016) * | To provide recommendations on the design of flashing lights at emergency exit portals for road tunnel emergency evacuation. | The experiment was carried out in a Cave Automatic Virtual Environment laboratory. A total of 96 participants eventually took part in the experiment (68 male and 28 female). Test participants' age ranged from 19 years to 64 years old (average = 25.15 years and standard deviation = 7.4 years). | Results show that green or white flashing lights perform better than blue lights. A flashing rate of 1 and 4 Hz performed better than a flashing rate of 0.25 Hz. A light emitting diode light source performed better than single and double strobe lights. The three layouts of the lights under consideration performed similarly. |
| (Cosma et al., 2016) | To investigate the impact of way-finding installations on the evacuation process. | The VR experiment presented in this article was carried out with Oculus Rift. The experiment reproduced an evacuation scenario from a smoke-filled railway tunnel. A between-group experimental design was chosen due to the objectives of the experiments. This includes a control group and two groups | The present study indicates that the use of way-finding lighting systems can be used to assist evacuees during route/exit choice. Another finding of the present study concerns the comparison of the two wayfinding lighting systems under consideration (alternate and continuous bright and dynamic green lights). Those systems performed similarly, that is, test participants do not show different responses to |
| | | navigating the tunnel with two different way-finding installations. Participants' coordinates over time were tracked during the entire duration of each experimental trial. At the end, a questionnaire was administered to the participants. | them. Results show that way-finding lighting installations have a positive impact on people evacuation safety. Given the limitations of the existing HMD technology in use, the results indicate that Oculus Rift can be used to qualitatively investigate human behavior in emergencies. Nevertheless, HMD devices give the possibility to collect data in a flexible, timely and cost-effective way. |
| (Zou et al., 2016) | To examine the feasibility of using a combination of subjective and objective measures, including an emotion scale and a physiological indicator, to assess the emotional responses of subjects in IVE-based evacuation experiments. | A total of 35 subjects were recruited to participate in the experiments. The subject wore SC data acquisition sensors by binding two electrodes around his/her index finger and middle finger and filled in a PANAS survey. Then the subject was asked to complete an evacuation task in IVE 1 and then IVE 2. The subject's physiological conditions, interactions with IVE, and first-person view from the HMD were monitored and recorded throughout the experiment using ErgoLAB platform. After each task, the subject was asked to fill up a PANAS survey. | The results showed that the PANAS emotion scale and the SC score were effective measures of emotional responses of the subjects to fire emergency IVEs. Assessment reported by both measures were generally in accordance with each other. The emotion scale was able to distinguish specific types of emotion responses elicited by the IVEs, especially scared, nervous and afraid, whereas the physiological indicators were more sensitive to subtle changes in the magnitude of emotional responses. When used together, the subjective and objective measures provided reasonable assessment of the sense of presence that subjects experience in emergency evacuation IVEs. |
| (Li et al., 2017) * | To provide an immersive and novel virtual reality training approach, designed to teach individuals how to survive earthquakes in common indoor environments. | The user experienced the simulation via the HTC VIVE in an empty space of 3m × 4m, the largest play area it allows. We recruited 96 participants, whose ages ranged from 20 to 30. They were undergraduate and graduate students from different majors. The participants were randomly divided into 4 groups of 24 people, with each group corresponding to a training condition described above. | Evaluation results show that our virtual reality training approach is effective, with the participants who are trained by our approach performing better, on average, than those trained by alternative approaches in terms of the capabilities to avoid physical damage and to detect potentially dangerous objects. |